\documentclass[aps,prb,amsmath,amssymb,11pt,a4paper,longbibliography,reprint,notitlepage,superscriptaddress]{revtex4-2} 

\usepackage{graphicx,graphics,epsfig}
\usepackage[colorlinks=true,allcolors=blue]{hyperref}
\usepackage[utf8]{inputenc}
\usepackage[english]{babel}
\usepackage{color}
\usepackage{SIunits}
\usepackage[normalem]{ulem} % to strike text

\begin{document}

\title{A chip-based superconducting magnetic trap for levitating superconducting microparticles}

\author{Martí Gutierrez Latorre}
\affiliation{Department of Microtechnology and Nanoscience (MC2), Chalmers University of Technology, SE-412 96 Gothenburg, Sweden}

\author{Achintya Paradkar}
\affiliation{Department of Microtechnology and Nanoscience (MC2), Chalmers University of Technology, SE-412 96 Gothenburg, Sweden}

\author{David Hambraeus}
\affiliation{Department of Microtechnology and Nanoscience (MC2), Chalmers University of Technology, SE-412 96 Gothenburg, Sweden}

\author{Gerard Higgins}
\affiliation{Department of Microtechnology and Nanoscience (MC2), Chalmers University of Technology, SE-412 96 Gothenburg, Sweden}
\affiliation{Institute for Quantum Optics and Quantum Information (IQOQI), Austrian Academy of Sciences, 1090 Vienna, Austria}

\author{Witlef Wieczorek}
\email{witlef.wieczorek@chalmers.se} 
\affiliation{Department of Microtechnology and Nanoscience (MC2), Chalmers University of Technology, SE-412 96 Gothenburg, Sweden}

\date{\today}

\begin{abstract}
    Magnetically-levitated superconducting microparticles have been recently proposed as a promising platform for performing quantum experiments with particles in the picogram regime. Here, we demonstrate the superconducting technology to achieve chip-based magnetic levitation of superconducting microparticles. We simulate and fabricate a chip-based magnetic trap capable of levitating superconducting particles with diameters  from 0.5$\,\mu$m to 200$\,\mu$m. The trap consists of two stacked silicon chips, each patterned with a planar multi-winding superconducting coil made of niobium. The two coils generate a magnetic field resembling a quadrupole near the trap center, in which we demonstrate trapping of a spherical 50\,$\mu$m diameter SnPb microparticle at temperatures of 4\,K and 40\,mK.
\end{abstract}


\maketitle %\maketitle must follow title, authors, abstract and \pacs

\section{Introduction}

Superconducting levitation is a well known phenomenon and allows levitation of objects of vastly different masses \cite{moon_superconducting_1994}. In the context of quantum experiments with macroscopic objects \cite{arndt_testing_2014}, superconducting levitation can enable a novel experimental platform combining ultra-low mechanical dissipation of levitated particles \cite{delic_cooling_2020,Leng_2021,tebbenjohannsQuantumControlNanoparticle2021,magriniRealtimeOptimalQuantum2021} with the capability to stably trap micrometer-sized objects  \cite{oriol,cirio_quantum_2012}. Theoretical proposals to realize macroscopic quantum superposition states \cite{Pino_2018}, as well as novel ultra-sensitive force and acceleration sensors \cite{Johnsson_2016,Prat-Camps_2017}, have recently been put forward that exploit these unique features.

Recent experiments have shown initial steps in this direction by levitating micro-magnets on top of superconductors \cite{niemetz_oscillating_2000,Wang_2019,timberlake_acceleration_2019,Vinante_2020,Gieseler_2020}, diamagnetic particles in strong magnetic fields \cite{Slezak_2018,zheng_room_2020,Leng_2021}, and superconducting microparticles in millimeter-sized superconducting magnetic traps \cite{Waarde_2016,hofer_2021}. Levitating a superconducting particle in a fully chip-based, microfabricated trap is advantageous as it enables high magnetic field gradients through miniaturization \cite{fortagh_magnetic_2007,dikovsky_superconducting_2009} and a straightforward integration of precisely positioned superconducting circuits for read-out and quantum control of the particle's motion \cite{oriol,cirio_quantum_2012,Johnsson_2016,Pino_2018}.

In our work, we  demonstrate a fully chip-based superconducting levitation platform. We discuss the design, simulation, and fabrication of our trap, and we use this trap in a proof-of-principle demonstration of levitating a 50$\,\mu$m-diameter superconducting particle. We base the design of our trap on recent theoretical work in which different magnetic trap architectures were analyzed \cite{oriol,cirio_quantum_2012,hofer_2019,marti, Navau_2021}.

\section{Chip-based superconducting magnetic trap}

\subsection{Levitation requirements}

In order to stably levitate a diamagnetic particle, first, a three-dimensional magnetic field minimum is required to confine it \cite{simon_diamagnetic_2000,marti}. The magnetic field distribution used to achieve such a confinement resembles that of an anti-Helmholtz coil configuration. At the levitation point, the magnetic force balances Earth's gravity. Second, to initially lift the particle off the substrate, the  magnetic lift force must also overcome adhesive forces between the particle and the substrate. From a simulation of our trap, we expect  a lift force of 19\,nN on a 50\,$\mu$m-diameter superconducting sphere in the Meissner state when we apply a current of 0.5\,A through the coils. This force greatly exceeds the particle's weight (0.5\,nN) and is, for certain particle-substrate interfaces, sufficient to overcome adhesive forces in our experiments. Note that these forces can be as large as 2\,$\mu$N for a 50\,$\mu$m metal sphere separated by 0.3\,nm from a flat surface \cite{vdWtheory}.

\subsection{Design and simulation of the magnetic trap}

\begin{figure*}[t!hbp]
\centerline{\includegraphics{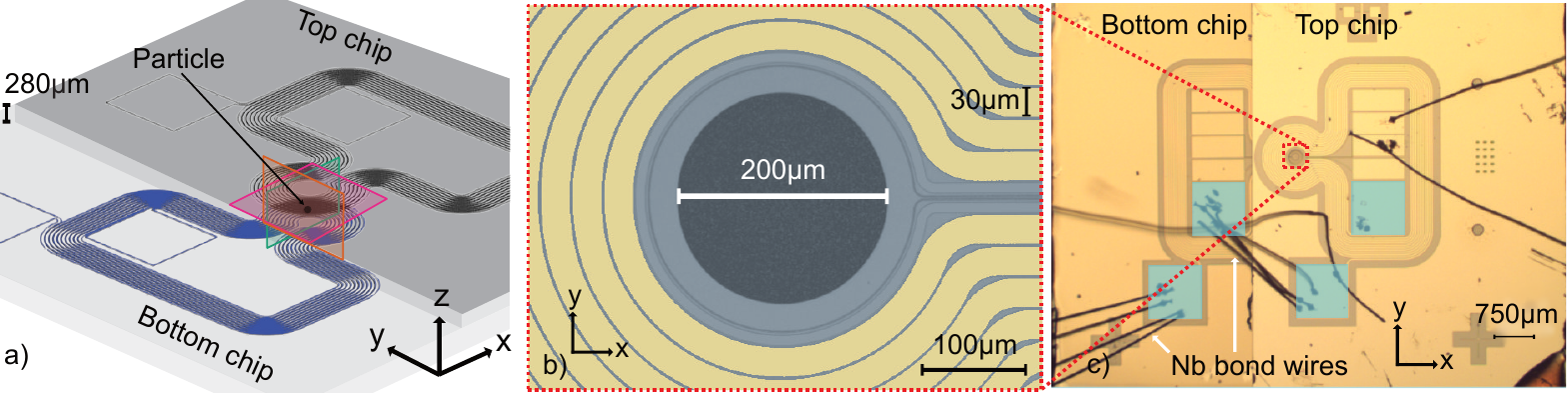}}
\caption{Chip-based magnetic trap. (a)~Schematic of the trap assembly, showing two 280\,$\mu$m thick silicon chips (top and bottom). The trap coils are colored in blue (bottom) and black (top), and the particle is shown between the coils. The colored rectangles indicate the orthogonal plane cuts used in Fig.~\ref{fig:Bfield}. (b)~False-color scanning electron microscope image of the top chip. The dark gray region is the hole where a particle is placed, the light gray region is the surface of the top silicon chip, and the yellow lines show the windings of the top coil. (c)~Assembled two-chip magnetic trap. The bond pads on the chip (light blue) are wire-bonded using $25\,\mu$m diameter niobium wires (black lines). The crosses are alignment markers used for etching the hole in panel (b).}
\label{fig:schematic}
\end{figure*}

\begin{figure*}[t!hbp]
\centerline{\includegraphics{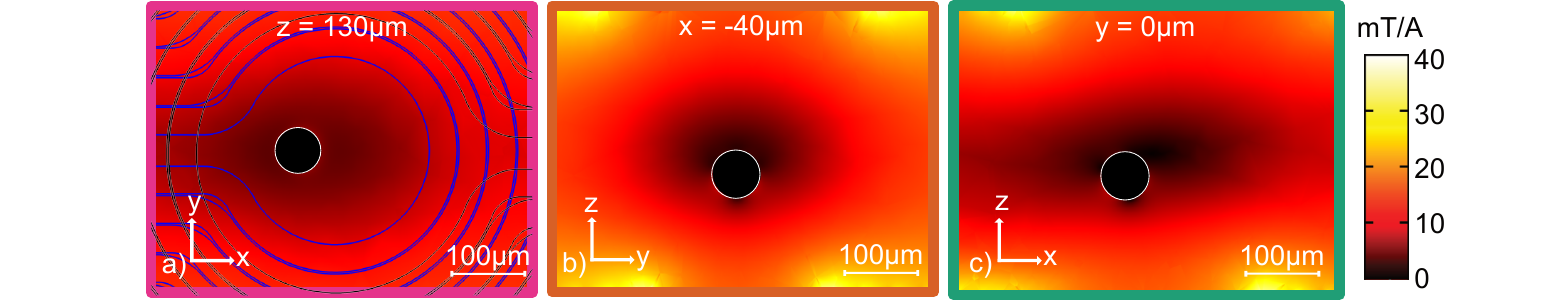}}
\caption{Simulated magnetic field strength in the trap scaled by the current in the coils, along the (a) $xy$, (b) $yz$, and (c) $xz$ planes. The center of the bottom coil defines the origin of the coordinate system. The displacement of each cut plane from the origin is given at the top of each panel. The equilibrium position of the levitated 50\,$\mu$m-diameter particle (black circle) is displaced from the origin along $z$ due to gravity and along $x$ due to the openings of the superconducting coils. Blue and black lines in (a) mark the projection of the bottom and top coil windings onto the $xy$ slice, respectively.}
\label{fig:Bfield}
\end{figure*}

The chip-based trap (see Fig.~\ref{fig:schematic}) consists of two multi-winding planar superconducting coils microfabricated on two silicon chips, which are stacked on top of each other. The coil separation is given by the 280\,$\mu$m thickness of one silicon chip. Each coil has ten windings with $2\,\mu$m spacing, and each winding has a rectangular cross-section with $30\,\mu$m width and $1\,\mu$m thickness. The coils are made of niobium and can carry a maximal current of 0.9$\,$A at 4\,K, limited by the critical current density of niobium (3$\cdot\text{10}^{10}\,\text{A/m}^2$ \cite{Hudson_1971}).
The coil dimensions are chosen as a trade-off between the magnetic field strength and the microfabrication yield: Narrower windings with smaller spacing can allow for a stronger magnetic field, due to the dependence of critical current density \cite{Hudson_1971} and current uniformity \cite{Ilin2010,talantsev_current_2018} on cross-section. However, millimeter-long  narrow wires frequently detach from the substrate or break during assembly when their height to width ratio exceeds about $0.5$.

To simulate the particle in the trap, we account for the three-dimensional geometry of the coils and the particle. We do this because the sample is not symmetric, thus, planar or rotationally symmetric two-dimensional models fail to capture its behavior, and also because the particle perturbs the trap field due to field expulsion \cite{lin_theoretical_2006,hofer_2019,marti}. We calculate the magnetic field distribution from finite element method simulations using COMSOL Multiphysics \cite{comsol}. The simulations operate in the static regime and are based on the A-V formulation of the Maxwell-London equations \cite{Campbell_2011}, for details see Ref.~\cite{marti}. The simulations yield the magnetic field (shown in Fig.~\ref{fig:Bfield}), magnetic force and supercurrents. We assume that the particle is in the Meissner state and model it via the Maxwell-London equations in A-V formulation \cite{marti}. This situation holds when the particle is cooled below its critical temperature in zero field, and if the field near the particle stays below the first critical field of the material that the particle is made of, as is the case in our experiment. The coils are made of a type-II superconductor; we capture this in the model assuming a nearly-perfectly diamagnetic material (relative permeability $0.001$) with a high electrical conductivity for the coils, as described in Ref.~\cite{talantsev_current_2018}. Hence, our model assumes that (i) the particle is in the Meissner state, and (ii) that there is no trapped flux in the levitated particle, inside the coil wire or inside the coil windings. Trapped flux would affect the levitation height, trap frequencies and mechanical dissipation.

Fig.~\ref{fig:Bfield} shows the trapping magnetic field, with a three-dimensional field minimum between the coils. The particle's equilibrium position is slightly shifted along the $z$ axis due to gravity, and along the $x$ axis due to the openings of the coils.

\begin{figure}[t!hbp]
\centerline{\includegraphics{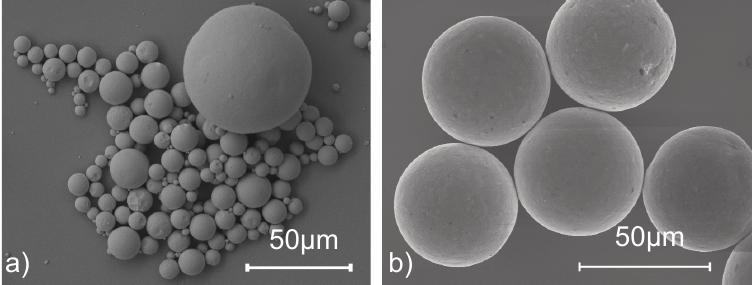}}
\caption{Scanning electron microscope images of superconducting microspheres used in the experiments. (a)~Lead microspheres produced by ultrasonic cavitation. (b)~Commercially-available tin-lead microspheres.}
\label{fig:particles}
\end{figure}

The simulated magnetic field allows us to estimate the trapped particle's centre-of-mass (COM) motional frequencies: We calculate the restoring force acting on the particle when it is shifted by a small distance from its equilibrium position \cite{marti}. In this way, we obtain motional frequencies for a 50\,$\mu$m particle,  $(\omega_x, \omega_y, \omega_z) = 2\pi\times(43\pm4, 64\pm19, 125\pm3)\,\text{Hz}$ when 0.5\,A current is passed through the coils (error is simulation uncertainty, see Ref.~\cite{marti}). These frequencies are much lower than the frequencies of typical micro- and nanomechanical resonators used in quantum experiments \cite{Aspelmeyer_Kippenberg_Marquardt_2014}. As a consequence, if the motional modes are in thermal states they will have around $10^9$ ($10^7$) phonons at 4\,K (10\,mK). Thus, in order for the levitated particle to be in its COM motional ground state additional cooling techniques, such as feedback \cite{tebbenjohannsQuantumControlNanoparticle2021,magriniRealtimeOptimalQuantum2021} or sideband cooling \cite{chan_laser_2011,teufel_sideband_2011}, are required.

The magnetic trap can be used to levitate particles of a huge range of diameters; the lower limit depends on the London penetration depth, $\lambda_L$, and the upper limit depends on the first critical field strength of the particle, $B_{c_1}$, or is given by the geometry of the trap. For example, SnPb particles with diameters from 500\,nm to 220\,$\mu$m could be levitated in our chip-based trap (assuming $\lambda_L=100\,\mathrm{nm}$ and  $B_{c_1}=40\,\mathrm{mT}$ \cite{Tsui2016,Lock1951}, when using a trap current of 0.5\,A).

\subsection{Fabrication}

The chip trap is fabricated using conventional microfabrication techniques: The substrate is a double-side-polished, 280$\,\mu$m-thick, two-inch undoped silicon wafer with (100) orientation. A $1\,\mu$m-thick Nb layer is deposited on the wafer by DC magnetron sputtering. The wafer is then diced into 7\,mm$\times$7\,mm chips with a Loadpoint Microace 3+ saw. The design for the magnetic trap is transferred to the Nb layer using electron beam lithography (Raith EBPG 5200) and reactive ion etching (Oxford Plasmalab 100). We etch a cylindrical hole into the top chip using the Bosch process with an STS ICP deep silicon etch system. The two chips are manually aligned using an optical microscope, and glued together using cryogenic GE-varnish. The top and bottom coils are electrically connected by wire bonding with a 25$\,\mu$m diameter niobium wire as described in Ref.~\cite{Jaszczuk_1991}; this ensures a superconducting connection between the coils at temperatures below 9\,K. Finally, a single particle is placed within the etched hole using a Signatone SE-10T tungsten tip attached to a xyz translation stage.

\begin{figure}[t!hbp]
\centerline{\includegraphics{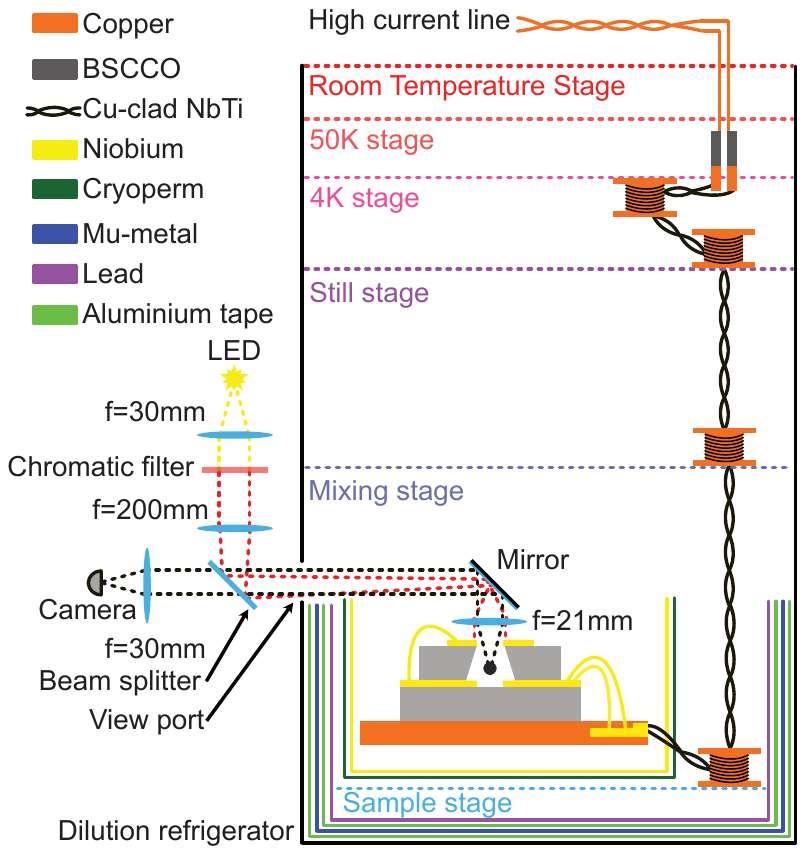}}
\caption{Schematic of the setup in the dilution refrigerator. The particle (black circle) levitates in the two-chip trap (not to scale) and is imaged using a custom-made microscope with K\"ohler illumination. The current supplied to the trap coils passes through superconducting wiring from the 50\,K stage down to the sample stage. The wires are thermalized at each stage using copper bobbins. The two-chip trap is glued to a copper holder that sits within a multi-layer magnetic shield.}
\label{fig:setup}
\end{figure}

\section{Superconducting microparticles}

We use spherical microparticles made of lead (Pb) or a tin-lead alloy (Sn63Pb37), shown in Fig.~\ref{fig:particles},
since lead and tin-lead have  critical temperatures higher than 4\,K (6.4\,K and 7.0\,K, respectively). Additionally, they have relatively high first critical field strengths (40\,mT and 80\,mT respectively \cite{Chanin1972}). We determined all these values, except for the critical field of lead, using AC magnetic susceptibility measurements.

The Pb microparticles are made in-house using ultrasonic cavitation \cite{Friedman_2010}, which yields near-spherical particles with a distribution of diameters between $0.5\,\mu$m and $200\,\mu$m. The tin-lead microspheres of 50$\,\mu$m diameter are commercially available (EasySpheres).

\section{Experimental setup}

The chip trap is placed in a dilution refrigerator, which has a viewport that provides optical access to the sample stage. We use a custom-made microscope employing K\"ohler illumination to monitor the particle on a CMOS camera.

\begin{figure*}[t!hbp]
\centerline{\includegraphics{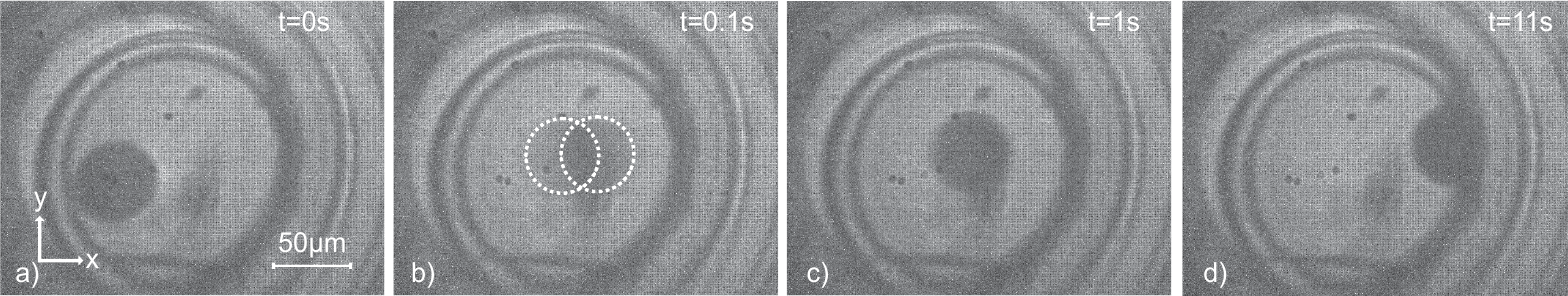}}
\caption{Top view of a single 50\,$\mu$m diameter microsphere in the chip trap before (a), during (b,c) and after (d) levitation. The time t is the time after current is applied to the trap coils. (a)~The particle rests on the substrate when no current is applied to the trap. (b)~The particle levitates and has a large oscillation amplitude when a current of 0.5\,A runs through the magnetic trap. The particle is seen as a blur because the oscillation period of the particle's motion is shorter than the camera's exposure time. The white dotted lines are a guide to the eye to identify the particle. (c)~The amplitude of the particle motion decreases until it is not resolved. The equilibrium position in (b) and (c) are the same. (d) Particle position after levitation has ceased.}
\label{fig:demo}
\end{figure*}

The trap current is supplied by an external current source, that passes through both pre-installed and custom-made wiring, as shown in Fig.~\ref{fig:setup}.
It is crucial to thermalize the wiring at each stage, to minimize the heat load at the sample stage. Pre-installed copper wires transport the current from room temperature to the 50\,K stage, then a pre-installed superconducting BSCCO tape transports the current to the 4\,K stage. At the 4\,K stage the BSCCO tape terminates in gold-plated copper connectors, to which we clamp a copper-clad, formvar insulated, 190\,$\mu$m twisted NbTi wire pair, which transports the current to the sample stage. At the sample stage, the NbTi wire is clamped to bulky Nb bond pads on the sample holder. At either end where the NbTi cable is clamped, the insulation and the cladding are removed to ensure a superconducting connection. The NbTi wire is wound and glued with GE-varnish around copper bobbins, which are attached to different stages for thermalization, as shown in Fig.~\ref{fig:setup}. The chip trap contains microfabricated Nb bond pads [see Fig.~\ref{fig:schematic}(c)], which are electrically connected to the bulky Nb bond pads on the sample holder via 25$\,\mu$m-diameter Nb wire bonds.

The sample is magnetically shielded (Fig.~\ref{fig:setup}) by two open-top baskets made of layers of superconducting and ferromagnetic material: the sample holder lies within a shield made of niobium and Cryoperm, and the entire sample stage lies within a shield made of lead, aluminium and mu-metal.

\section{Demonstration of magnetic levitation}\label{Demonstration}

\begin{figure}[b!thbp]
\centerline{\includegraphics{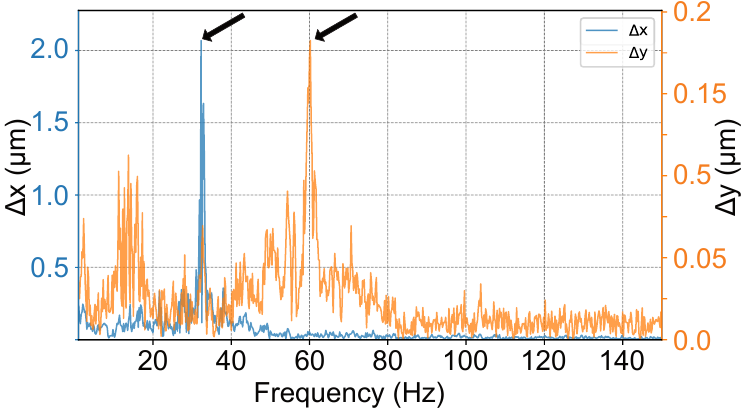}}
\caption{Frequency spectrum of the particle's motion at a trap current of 0.5\,A, along the $x$ and $y$ directions, extracted from analyzing the camera videos.}
\label{fig:motion}
\end{figure}

In Fig.~\ref{fig:demo} we show images of a SnPb particle in the magnetic trap at a temperature of 4\,K (see also video material \cite{zenodovideos}). When we apply sufficient current to the trap, the particle levitates. This is evidenced by a sudden change of the particle position from its initial rest position [Fig.~\ref{fig:demo}(a)], additionally the particle becomes blurred because of its oscillatory motion about the trap center [Fig.~\ref{fig:demo}(b)]. The motional amplitude is strongly damped after 1\,s [Fig.~\ref{fig:demo}(c)]. After about 10\,s the continuous illumination causes superconductivity to break and as a result the particle falls down [Fig.~\ref{fig:demo}(d)].  When we do not shine light on the particle, we can levitate it for several days. We observe similar levitation results when using pure lead microparticles and also when the trap is operated at 40\,mK (see also video material \cite{zenodovideos}).

We extract the particle's motional frequencies from an analysis of the levitation video. To this end, we determine the time-dependent particle center by fitting an ellipse to the particle images and identify the center position of the ellipse with the position of the particle in the $xy$ plane. A Fourier transform of this time-dependent position yields the data in Fig.~\ref{fig:motion}. We can clearly identify two dominant frequency components and attribute these to the COM motional modes along the $x$ and $y$ direction, $(\omega_x,\omega_y)=2\pi\times(32\pm1.6,60\pm1.8)\,$Hz, respectively. The frequencies are in good agreement with the simulation results of $(\omega_x, \omega_y) = 2\pi\times(43\pm4, 64\pm19)\,\text{Hz}$. We attribute the small discrepancy to the spatial misalignment between the bottom and top coils, as is visible in Fig. \ref{fig:demo}.

\section{Conclusion and outlook}

We have presented the design, simulation and fabrication of a chip-based superconducting magnetic trap for levitating superconducting microparticles of diameters between 0.5$\,\mu$m and 200$\,\mu$m. We have demonstrated stable levitation of 50\,$\mu$m-diameter superconducting particles in this trap at temperatures of 4\,K and 40\,mK. In the future, we will lower the particle's motional dissipation rate by using a DC-SQUID magnetometer to detect the particle motion non-invasively and by improving the magnetic field shielding. These are crucial steps to eventually bring the particle's COM motion into the quantum regime. Note that in an independent recent experiment, chip-based superconducting magnetic levitation of a superconducting microparticle has also been realized \cite{trupke2021}.

\begin{acknowledgments}
We thank Thilo Bauch and Avgust Yurgens for insightful discussions. Further, we appreciate fruitful discussions with Joachim Hofer, Philip Schmidt, Stefan Minniberger, Michael Trupke, Markus Aspelmeyer and the other members of the EU Horizon 2020 project MaQSens. We are thankful for initial microfabrication support from David Niepce. We acknowledge funding from the Knut and Alice Wallenberg foundation through a Wallenberg Academy fellowship (W.W.), the Wallenberg Center for Quantum Technology (WACQT, A.P.) and support from Chalmers Excellence Initiative Nano. G.H.~acknowledges support from the Swedish Research Council (grant no.~2020-00381). Sample fabrication was performed in the Myfab Nanofabrication Laboratory at Chalmers. Simulations were performed on resources provided by the Swedish National Infrastructure for Computing (SNIC) at Tetralith, Linköping University, partially funded by the Swedish Research Council through Grant No.~2018-05973.
\end{acknowledgments}

%\clearpage

\bibliography{bib_sctrap}

\end{document}